\documentclass[]{spie}  
\usepackage[]{graphicx}
\usepackage{aas_macros}
\usepackage{color}

\title{The FORCE mission : \\
Science aim and instrument parameter for broadband X-ray imaging spectroscopy with good angular resolution}

\author{Kazuhiro Nakazawa\supit{a}, Koji Mori\supit{b}, 
Takeshi G. Tsuru\supit{c},
Yoshihiro Ueda\supit{d},
Hisamitsu Awaki\supit{e},
Yasushi Fukazawa\supit{f},
Manabu Ishida\supit{g},
Hironori Matsumoto\supit{h},
Hiroshi Murakami\supit{i},
Takashi Okajima\supit{j},
Tadayuki Takahashi\supit{k},
Hiroshi Tsunemi\supit{h},
and William W. Zhang\supit{j}
\skiplinehalf
\supit{a}Kobayashi-Maskawa Institute, Nagoya University, Furo-cho, Chikusa-ku, Nagoya, 464-8602, Japan \\
\supit{b}Department of Applied Physics and Electronic Engineering, University of Miyazaki, Miyazaki, 889-2192, Japan\\
\supit{c}Department of Physics, Kyoto University, Kyoto, 606-8502, Japan\\
\supit{d}Department of Astronomy, Kyoto University, Kyoto, 606-8502, Japan\\
\supit{e}Department of Physics, Ehime University, Ehime, 790-8577, Japan\\
\supit{f}Department of Physical Science, Hiroshima University, Hiroshima, 739-8526, Japan\\
\supit{g}Institute of Space and Astronautical Science, JAXA, Chuo-ku, Sagamihara, 252-5210, Japan \\
\supit{h}Department of Earth and Space Science, Osaka University, Osaka,  560-0043, Japan\\
\supit{i}Department of Information Science, Faculty of Liberal Arts, Tohoku Gakuin University, Miyagi 981-3193, Japan\\
\supit{j}NASA/Goddard Space Flight Center, MD 20771, USA\\
\supit{k} Kavli IPMU, UTIAS, The University of Tokyo, Kashiwa, Chiba, 277-8583, Japan\\
}

\authorinfo{Further author information: (Send correspondence to K. N.)\\K. N.: E-mail: nakazawa@u.phys.nagoya-u.ac.jp, Telephone: 81 52 788 6268
}

\begin{document} 
  \maketitle 

\begin{abstract}
FORCE is a 1.2 tonnes small mission dedicated for wide-band fine-imaging x-ray observation. It covers from 1 to 80 keV with a good angular resolution of $15''$ half-power-diameter. It is proposed to be launched around mid-2020s and designed to reach a limiting sensitivity as good as $F_X (10-40~{\rm keV}) = 3 \times 10^{-15}$~erg cm$^{-2}$ s$^{-1}$ keV$^{-1}$ within 1~Ms. This number is one order of magnitude better than current best one. With its high-sensitivity wide-band coverage, FORCE will probe the new science field of ``missing BHs'', searching for families of black holes of which populations and evolutions are not well known. Other point-source and diffuse-source sciences are also considered. FORCE will also provide the ``hard x-ray coverage''  to forthcoming large soft x-ray observatories.
\end{abstract}



\section{Introduction}
\label{sect:intro}  

X-ray imaging spectroscopy with wide band coverage provides vital information on high energy phenomena in the universe. With better angular resolution, dimmer point sources can be detected, and detailed structure of diffuse sources can be identified. With wider energy coverage, emission mechanisms of x-rays and its physical parameters can be well estimated, and therefore the nature of the source becomes clearer. 
For example, x-ray spectra of typical nearby active galactic nuclei (AGN) composes of; a soft-excess component  and absorption features at $<1$~keV , power-law (PL) like component around 2--10~keV, the Iron fluorescence K-line emission structure around $\sim6$~keV from reflected components, and the Compton hump of it around $\sim 30$~keV, coupled with the PL spectral cutoff around $\sim 100$~keV (e.g. \cite{2014arXiv1412.1177R}). To understand the mass, density, temperature and geometry of materials surrounding the central engine, all of these information are needed.

The {\it Hitomi} x-ray observatory\cite{hitomi2018JATIS} launched in 2016 was characterized by the superb energy resolution $\sim 5$~eV in full-width at half maximum (FWHM) at 0.5--12~keV with the Soft X-ray Spectrometer (SXS) instrument\cite{Mitsuda2012}, but it also aimed at simultaneous wide-band spectral coverage using the Soft X-ray Imager (SXI)\cite{10.1117/1.JATIS.4.1.011211}, an x-ray CCD imager coupled with the Soft X-ray Telescope (SXT), and the Hard X-ray Imager (HXI)\cite{10.1117/1.JATIS.4.2.021410} coupled with the Hard X-ray Telescope (HXT)\cite{Awaki:14} providing imaging spectroscopy in the 5--80~keV range, and the Soft Gamma-ray Detector (SGD) covering up to 600~keV\cite{10.1117/1.JATIS.4.2.021411}. Because {\it Hitomi} is not functional now, the current best approach to obtain wide-band high-sensitivity imaging spectroscopic data is combining {\it NuSTAR}\cite{2013ApJ...770..103H}, with its $\sim 1'$ half-power diameter (HPD) imaging spectroscopy in 3--80 keV, coupled with soft x-ray observatories, such as {\it Chandra} and {\it XMM-Newton}.

\begin{figure}
\begin{center}
\includegraphics[height=8cm]{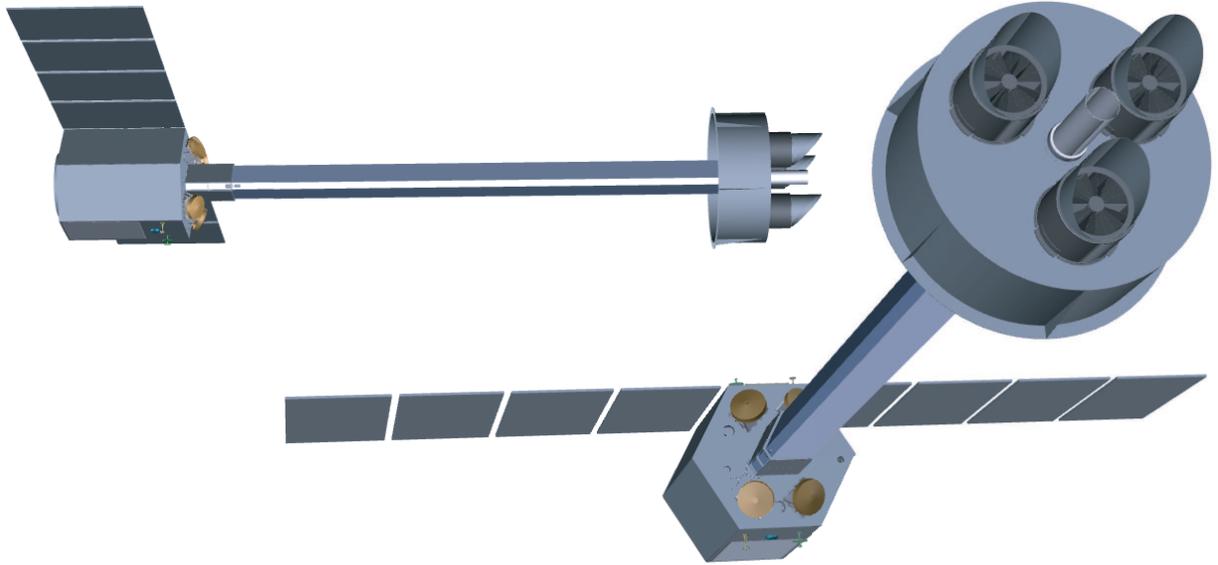}  
\end{center}
\caption 
{ \label{fig:force} Rough 3D drawing of the FORCE mission. Shown is the configuration in orbit, after deploying $\sim10$~m of extensible optical bench.} 
\end{figure}

FORCE ({\it Focusing On Relativistic universe and Cosmic Evolution}) is a mission we are proposing to provide a wide-band imaging spectroscopy in 1--80~keV with a good angular resolution of $\sim10''$ HPD\cite{2016SPIE.9905E..1OM}. The mission is proposed for launch around mid-2020s, to provide an order of magnitude better sensitivity than {\it NuSTAR} or {\it Hitomi}/HXI-HXT. The satellite is relatively small, with total weight of $\sim 1.2$~tonnes, to fit within the envelope of JAXA Epsilon rocket. As shown in Figure~\ref{fig:force}, the long focal length important for hard x-ray optics is obtained by using an extensible optical bench (EOB). In this paper, we will shortly summarize the science aims (\S \ref{sect:science}), requirements and on-board instrument's design (\S \ref{sect:technical}), and its synergy with the soft x-ray missions, such as XARM\cite{2018SPIE_XARM}, ATHENA and/or Lynx (\S \ref{sect:discussion}).

\section{Science aims of the FORCE mission}
\label{sect:science}

The primary science goal of FORCE is to identify and estimate the number of ``missing black holes''; families of black holes (BHs) of which populations and evolutions are not well known. Namely, highly-obscured super-massive BHs (SMBHs) with a mass of  $M_{\rm BH} \sim 10^{5}$--$10^{11}$~$M_{\odot}$ ($M_{\odot}$ is a solar mass), yet unidentified intermediate-mass BHs (IMBHs) with $M_{\rm BH} \sim 10^{2}$--$10^{4}$, and isolated stellar-mass BHs with $M_{\rm BH} \sim 10$--$10^{2}$   which are estimated to be numerous. Number density of these BHs are key to deepen the understanding of evolution paths of the universe. Other point-source will be investigated as well, such as neutron star (NS) low-mass x-ray binaries (LMXBs) and high-mass X-ray binaries,   radio pulsars, magnetors, and cataclysmic variables (white dwarf binaries). In addition, adoption of a detector with low background (BGD) level combined with a good angular resolution provides ideal probe to study diffuse emission from hottest and non-thermal plasmas.

\subsection{Search for missing BHs}

\subsubsection{Highly obscured AGN}

Cosmic X-ray Background (CXB) is a summation of x-ray emission from numerous AGNs, with a variety of luminosity,  a variety of absorption column density ($N_{\rm H}$), located in a variety of redshift (e.g. \cite{2014ApJ...786..104U}). AGN emission originates from mass accretion to SMBHs, and therefore is related to their growth. To understand the AGN evolution history in the universe, multiwavelength surveys including X-ray ones are extensively performed.  

A wide-band x-ray spectroscopy study by {\it Suzaku} and {\it Swift} in the 0.5--40~keV band has revealed there are also a variety in the covering fraction of absorbing materials around the SMBH in AGNs \cite{2007ApJ...664L..79U,2017Natur.549..488R}. This fraction can be measured by comparing the flux of scattered soft x-ray component to that of the primary PL emission strongly absorbed and only seen in the hard x-ray band. In other words, only with wide-band x-ray spectra. AGNs with large covering fraction of matter are called ``buried AGNs''. Kawamuro et al. (2016)\cite{0067-0049-225-1-14} compared the hard x-ray (10--50~keV) luminosity of  obscured AGNs (both buried and non-buried) to that of the MIR emission lines from highly ionized ions in the narrow line region, namely the [O IV]~26$\mu$m line. They found that buried AGNs with similar [O IV]~26$\mu$m emission line intensity have about an order of magnitude larger x-ray luminosity than non-buried AGNs. 
A good example is a galaxy NGC~4945, which was dim in the  [Ne V] 14.3 $\mu$m line (also considered to be related to AGN activity) but has very high hard x-ray luminosity and now identified as hosting an AGN\cite{1993ApJ...409..155I,2011A&A...533A..56P}. 
%
These buried AGNs are known to be ubiquitous in galaxies with high star-formation rates (e.g. \cite{2017ApJ...835..179O}), and hence are key objects to understand the origin of the galaxy-SMBH co-evolution. X-ray surveys are indispensable to search for these populations in the most complete manner. 

\begin{figure}
\begin{center}
\includegraphics[height=7.0cm]{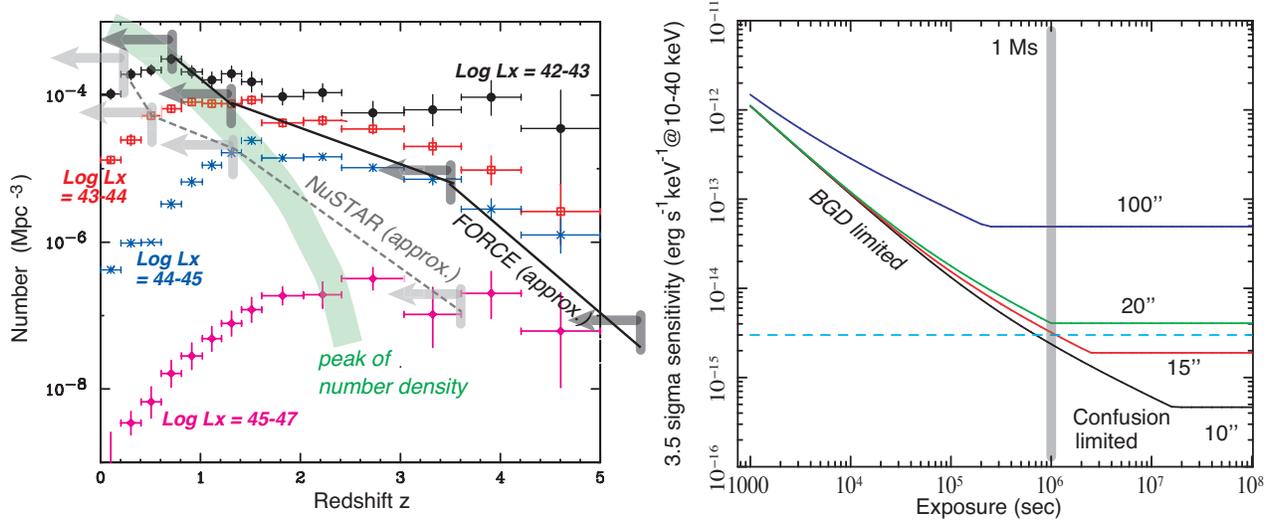}  
\end{center}
\caption 
{ \label{fig:agn}
{\it (left)}~ Number density of Compton-thin AGNs with various luminosity, from Ueda et al. (2014)\cite{2014ApJ...786..104U}. Shown in gray is the sources observable with NuSTAR (thick light gray) and FORCE (thick dark gray). Number density of Compton-thick AGNs is not well known, which is the finding space for FORCE. {\it (right)}~ Plot of limiting sensitivity as a function of exposure. Four cases of imaging sharpness are presented; $100''$ HPD representing {\it Hitomi}/HXI-HXT, and $20''$, $15''$, and $10''$ for comparison. Modified from Mori et al. (2016)\cite{2016SPIE.9905E..1OM}. See text for detail.} 
\end{figure}

Using data from existing x-ray observatories, Ueda et al. (2014)\cite{2014ApJ...786..104U} derived the number density of Compton-thin AGNs ($N_{\rm H} < 10^{-24}$~cm$^{-2}$) sorted with  luminosity, as shown in the left panel of Figure~\ref{fig:agn}. They found a striking result that, more luminous AGN has a peak in their number density in the past. In general, AGN luminosity is considered to be related to the SMBH mass, in view of Eddington luminosity. Combining these two aspects, their finding suggest that SMBHs with larger mass have evolved earlier, and those with smaller mass evolve later. This is named ``downsizing'' and remains one of the big mysteries in the BH evolution of the universe. 
However, the ``down sizing'' is not observationally confirmed in Compton-thick AGNs, in which AGNs with  higher mass accretion rate are considered to reside. To address this question, high sensitivity in hard x-ray band is required. 

In Figure~\ref{fig:agn}-left, we overlaid the approximate sensitivity range of {\it NuSTAR}, assuming that Compton-thick AGN has similar evolution and number density as Compton-thin ones. It shows that we need an order of magnitude better sensitivity to actually follow the evolution peak. From here, we derived the sensitivity requirement in 10--40~keV band of FORCE to be $F_X (10-40~{\rm keV}) = 3 \times 10^{-15}$~erg cm$^{-2}$ s$^{-1}$ keV$^{-1}$.  With this sensitivity limit, FORCE will be able to resolve $\sim80$\% of the CXB emission into point sources, and hence identify the origins of the CXB in its 30~keV flux peak.  

In the right panel of Figure~\ref{fig:agn}, we presented the sensitivity as a function of exposure and angular resolution. With smaller mirror HPD, the BGD contaminating in the source integration region decreases, making the statistical sensitivity limit better. But, ultimately, the angular resolution itself will define the limiting sensitivity in view of confusion.  From the plot, an angular resolution of better than $15''$ HPD, or somewhere around $10''$ HPD, is needed to achieve the required sensitivity of $3 \times 10^{-15}$~erg cm$^{-2}$ s$^{-1}$ keV$^{-1}$. If the effective area of the mission is  $350$~cm$^2$ at 30 keV (similar to {\it NuSTAR} and {\it Hitomi}/HXI-HXT), reaching the sensitivity will require 1~Ms of observation. FORCE will be able to evaluate the luminosity-sorted evolution peaks and these   results will provide unprecedented clue to understand AGN evolution, including both Compton-thin and thick AGNs. 

\subsubsection{Intermediate and stellar mass BHs}

The detection of gravitational wave from binary BH merger forming a BH with $\sim 62~M_{\odot}$\cite{2016PhRvL.116f1102A}   motivated   us to revisit intermediate-mass BHs with $M_{\rm BH} \sim 10^2-10^{4}~M_{\odot}$, which connects between SMBHs and stellar-mass BHs. Ultra-luminous and hyper-luminous X-ray sources (ULXs and HLXs) are x-ray point sources with  luminosity   largely exceeding the Eddington limit of a NS with as mass of $M \sim 1.4~M_{\odot}$ (e.g. \cite{2017ARA&A..55..303K} and \cite{2009Natur.460...73F}). Even though existence of  ULXs   was known from late 1970s, their central engine are not well known yet. Recent detections of clear pulsation from a few ULXs (e.g. \cite{2014Natur.514..202B}) have shown that, at least a certain fraction of ULXs are NSs with super-Eddington luminosity. In parallel, a theoretical model of a mass-accreting BH to reproduce the super-Eddington luminosity, and  the characteristic spectral shapes of ULXs, has been proposed\cite{2012ApJ...752...18K}. It is   now also expanded to super-critical accretion on to a NS\cite{0004-637X-853-1-45}. High-sensitivity wide-band spectral observation study of ULX/HLX coupled with progress in theoretical study will provide good tool to understand the central engine of these sources. Because these sources are located within neighboring galaxies, angular resolution to avoid confusion is critically important, as shown by the {\it NuSTAR} observations (e.g. \cite{2014ApJ...797...79W}).

Amount of stellar mass BHs reflects the initial stellar mass function and evolution paths of high-mass star in our galaxy. From simple estimation, we  expect 1--10 millions of stellar-mass BHs in our galaxy, and majority of them are considered to be isolated, and hence dim in all frequency bands. To date, we know only 20--30 of stellar mass BH, most of which are detected as x-ray flareup of binary systems. Survey for quiescent BH-LMXB (e.g. \cite{2018Natur.556...70H,2016ApJ...825..132H}), and isolate BHs\cite{2018MNRAS.475.1251M,2018MNRAS.477..791T} interacting with molecular clouds are important approach to address this question. What  is also interesting is the detection of high-velocity compact molecular clouds in the radio band, which is proposed to be a remnant of interaction between IMBH and these clouds  (e.g. \cite{2016ApJ...816L...7O}). X-ray emission from these sources are considered to be dim (e.g. \cite{2016SPIE.9905E..1OM}). Because they also reside in densely populated galactic ridge regions, identification from other x-ray sources is needed. Wide-band x-ray spectra is suggested to be distinctly non-thermal and hard   if the central engine is a BH\cite{2006ARA&A..44...49R}, and FORCE can provide key information here.

\subsubsection{other point source sciences}

Among many other candidate science targets on point source observation with FORCE, we would like to point out two different aspects. First of all, wide band coverage from soft x-rays (down to $\sim 1$~keV) to hard x-rays (up to $\sim 80$~keV) provides diagnosing power to understand the physical nature of the source. It provides a tool to identify how many components are contributing in the AGN x-ray spectra (e.g. \cite{2016ApJ...828...78N}), and identifying the relation between enigmatic hard components and soft component in magnetars\cite{2017ApJS..231....8E}. Also, with the required sensitivity, FORCE can detect the x-ray afterglow\cite{2017Natur.551...71T} of the NS-NS merger event GW170817\cite{PhysRevLett.119.161101}, although the physical aspects of such observations is not  clear at this moment. With the improving gravitational wave observations, and large optical transient survey such as LSST coming soon, wide-band and high-sensitivity x-ray follow up observation capability provides strong tool.

\subsection{impact on diffuse source science}

The low and stable detector BGD combined with modest effective area 
of  FORCE  also provides ideal tool to observe diffuse hard x-ray emission. 
Notably, the good angular resolution to resolve most of the CXB point sources means the fluctuation of CXB flux caused by the number distribution of these point sources (called ``CXB fluctuation'') can be significantly reduced. In other words, ``almost CXB free observation'' is possible and this will greatly improve the accuracy of dim diffuse source measurements. 
For example, the post-shock electron temperature of binary cluster merger CIZA J1358.9-4750\cite{2015PASJ...67...71K} is as high as $kT \sim 12$~keV, and hence cannot be well determined with soft x-ray observatories such as {\it Chandra} and {\it XMM-Newton}. {\it NuSTAR} observation of the Bullet Cluster showed that hard x-ray imaging spectroscopy supplemented by soft x-ray one can well determine the electron temperature\cite{2014ApJ...792...48W}. At the same time, they showed that low and stable detector BGD is critical to identify the hottest component or non-thermal component. FORCE will greatly improve the sensitivity for hard diffuse source, by providing $\sim 5$~times better angular resolution and $\sim 4$~times lower BGD (see \S \ref{sect:technical}).

\section{Technical aspects of the FORCE mission}
\label{sect:technical}

\subsection{technical requirements and overall instrument design}

The FORCE mission is designed to carry three sets of high-resolution hard x-ray telescopes (HXT) and wide-band hybrid x-ray imagers (WHXI). The HXTs has a focal length of 10~m, and an EOB is adopted to position the imagers at their foci in orbit (Figure~\ref{fig:force}). The top level requirement is to be able to detect a source with a flux as low as $F_{\rm X} =3 \times 10^{-15}$~erg cm$^{-2}$ s$^{-1}$ keV$^{-1}$ within 1~Ms. All the other flown-down requirements are listed in Table~\ref{tab:spec} (see also \cite{2016SPIE.9905E..1OM}).

\begin{table}[ht]
\caption{Requirement list for FORCE.} 
\label{tab:spec}
\begin{center}       
\begin{tabular}{ll} 
\hline\hline
Angular resolution (HPD) & $<15''$\\
Multi-layer Coating		& Pt/C\\
Field of view (50\% response) at 30 keV & $>7'\times7'$\\
Effective Area at 30 keV & 370 cm$^2$\\
Energy range & 1--80~keV\\
Energy resolution (FWHM) at 6 keV & $<300$~eV \\
~~~~~~~~~~~~~~~~at 60 keV & $<1$~keV\\
Background & comparable to those of {\it Hitomi}/HXI-HXT \\
	& (1--3$\times10^{-4}$~cts s$^{-1}$ cm$^{-2}$ keV$^{-1}$)\\
Timing resolution	& several $\times 10$~$\mu$s\\
Working temperature & $-20\pm1$~$^{\circ}$C\\
\hline\hline
\end{tabular}
\end{center}
\end{table}

\subsection{the hard x-ray telescope (HXT)}

The technology is based on light-weight, polished Si mirror developed by the team at NASA/GSFC\cite{2014SPIE.9144E..15Z} (see also W. W Zhang et al. this conference). Recently, the team fabricated a single pair of mirror,   and has shown $\sim 2''.2$ HPD angular resolution with $\sim4$~keV x-rays. Based on the technology, the HXT of FORCE is to be made of    the Si substrate with 400~$\mu$m thickness, and has a diameter of 440~mm.

As the reflecting surface of the mirror, we plan to deposit a super-mirror multi-layer coating made of Pt/C to enhance the photon reflectivity in the hard x-ray band, above $\sim15$~keV. The base-line technology is those used for {\it Hitomi}/HXT. Super-mirror coating on a fine angular resolution thin mirror substrate can cause two types of degradation in its angular resolution. The first one is the roughness of the coating, which scatters x-rays and reduces the reflectivity. Using a test Si substrate deposited with Pt/C multi-layer, we are verifying the image sharpness using the 30~keV beam-line at SPring-8.   The second one is the stress between the Si substrate and multi-layer coating, which will modify the shape of the substrate. Currently technologies to counter this stress is under development. Because the optics is the key to the success of this mission, we are also sharing some informations with other hard x-ray optics developers, to control the development risk.

\subsection{the wide-band hybrid x-ray imager}

\begin{figure}[htbp]
\begin{center}
\includegraphics[height=9.0cm]{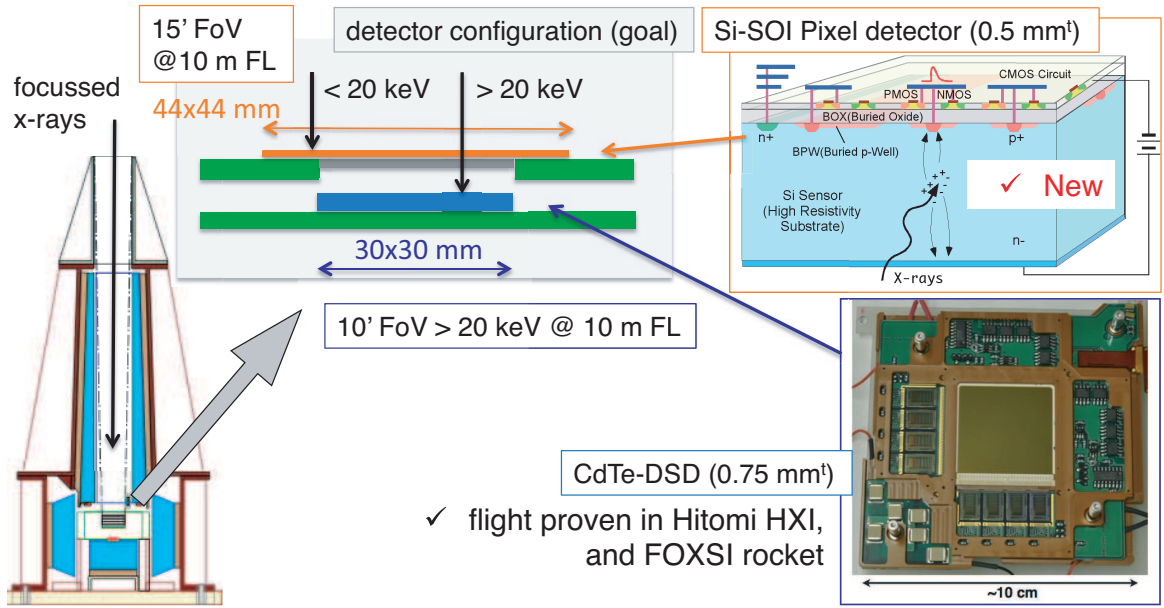}  
\end{center}
\caption 
{ \label{fig:whxi} Concept drawing of the WHXI detector.} 
\end{figure}

Schematic view of the WHXI system is presented in Figure~\ref{fig:whxi}. The overall detector design is based on that of {\it Hitomi}/HXI\cite{10.1117/1.JATIS.4.2.021410}. The HXI was made of the imager part and the active-shield part which was almost completely surrounding the former. Active shield was made of 1--3~cm thick BGO scintillator crystals to perform anti-coincidence. The imager was made of four layers of double-sided Si strip detectors (DSSDs), and a CdTe double-sided detector (CdTe-DSD). The five layers of imagers provided ``depth sensing'' so that optimized ``thickness'' can be selected. For example, below $\sim 12$~keV the first DSSD absorbs all photons, while any signal detected in other layers is BGD. Combined with deep active shielding, the HXI was designed to provide the lowest detector BGD ever achieved in the hard x-ray band.
 
When the tragedy caused the loss of the {\it Hitomi} mission, the HXI was fully operational in orbit and just started the scientific observations. Based on the in-orbit performance, it was shown that the detector BGD level was as low as 1--3$\times10^{-4}$~cts s$^{-1}$ cm$^{-2}$ keV$^{-1}$ at 5--80~keV. This number is $\sim4$ times lower than those of {\it NuSTAR} FPM especially around 5--30~keV\cite{10.1117/1.JATIS.4.2.021410}. By detailed analysis of the BGD properties, we also found that the component originating from $\sim 100$~keV electrons dominating below $\sim 30$~keV, can be reduced significantly if additional shielding were to be added. In other words, BGD level can be reduced further by 30--50\% (see Hagino K. et al. this conference).

The largest modification of the WHXI from {\it Hitomi}/HXI is the adoption of Si-SOIPIX imager in place of the DSSDs, to expand the soft x-ray coverage from 5~keV down to 1~keV. The Si-SOIPIX imager is a low-noise Si-CMOS imager specifically designed to have a self triggering capability; detect the x-ray photon absorbed and trigger the data acquisition sequence by itself. It hence has a time tagging accuracy of $<10~\mu$s. With this functionality, the imager can perform anti-coincidence while also covering down to 1~keV. It also provides high throughput rate (as high as 2~kHz designed). In the current design, the Si-SOIPIX imager has an area of $44\times44$~mm$^2$, covering $15' \times15'$ FoV. Below the Si imager, a CdTe-DSD is located to cover hard x-ray photons from $\sim 20$ to 80~keV, heritage of {\it Hitomi}/HXI. It has a size of $30\times30$~mm$^2$, covering $10' \times10'$ FoV.  To minimize the de-focusing effect, the two imagers are located within 4~mm distance. See Tsuru G. T. et al. in this conference for the most updated status of the SOIPIX imager. 

\subsection{mission design progress}

Currently, the FORCE mission is in concept phase, and we are planning to propose this mission to be launched at mid-2020s. Because it needs long focal length and good attitude determination, some key technological challenges must be solved. Because the satellite is very long, gravitational tidal force is non-negligible. The attitude control system needs to be carefully designed so that the reaction wheels have enough power, and magnetic torquers are powerful enough to  effectively damp the angular momentum accumulated in orbit. The EOB is planned to be an enlarged version of those used in {\it Hitomi}. Attitude determination is provided by the star tracker at the top of the satellite, co-alligned to the observational axis. Alignment monitoring system to measure the distortion of the optical bench is also adopted.

\section{Importance of hard x-ray imaging spectroscopy mission in 2030s}
\label{sect:discussion}

\begin{figure}
\begin{center}
\includegraphics[height=7cm]{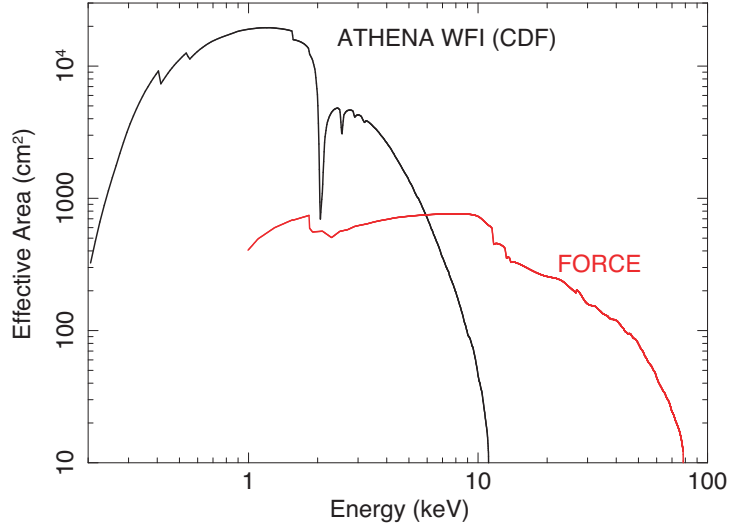}  
\end{center}
\caption 
{ \label{fig:EA} Effective area comparison between ATHENA and FORCE.} 
\end{figure}

In early 2030s, ESA will be operating the large soft x-ray observatory {\it ATHENA} in orbit. NASA is also starting a study on similarly large soft x-ray mission {\it Lynx}. On the other hand, contemporary hard x-ray coverage is currently lacking. In Figure~\ref{fig:EA}, we compared the effective area of {\it ATHENA} and FORCE. It is clear that the former mission has a huge x-ray collecting area around 1~keV, exceeding 1~m$^2$. On the other hand, the area rapidly drops above $\sim 3$~keV and even crosses with FORCE at around 6~keV. This is simply because the x-ray optics of {\it ATHENA} is optimized for $\sim 1$~keV, resulted in a $\sim 3$~m diameter large single x-ray optics with a focal length of 12~m. The FORCE's optics is optimized around 30~keV. It has similar focal length of 10~m, but carries three identical mirrors with a diameter of 44~cm. Because layers outside this diameter cannot reflect 30~keV photons due to too large grazing angle even with the multi-layer coatings, there is no merit in making the optics' diameter larger.

The resultant effective area comparison is interesting. The effective area at $\sim 6$ keV of {\it ATHENA} is $\sim 3$ times larger than that of FORCE at 20~keV. This ratio is very similar to that of the sum of four soft x-ray imagers onboard {\it Suzaku} at 6~keV and that of the HXD-PIN at 20~keV.  The wide-band capability of {\it Suzaku} was the vital characteristics on providing analyzing power to resolve a few components on AGN emission\cite{2016ApJ...828...78N} as well as the hard component in magnetors\cite{2017ApJS..231....8E}. Combination of {\it ATHENA} and FORCE will bring new science, with much improved sensitivity in both the soft and hard x-ray bands.

\section{Conclusion}

FORCE is a mission for wide-band fine-imaging x-ray observation, covering from 1 to 80 keV, with $15''$ HPD good angular resolution. It is a 1.2-tonnes small mission proposed to be launched around mid-2020s. With the high angular-resolution Si mirror with super-mirror coating combined with the lowest and stable BGD of the WHXI, it is designed to provide a limiting sensitivity as good as $F_X (10-40~{\rm keV}) = 3 \times 10^{-15}$~erg cm$^{-2}$ s$^{-1}$ keV$^{-1}$ within 1~Ms. With its high-sensitivity wide-band coverage, FORCE will probe the new science from pursuing ``missing BHs'' and other point-source and diffuse-source sciences as well as providing ``additional'' band coverage to forthcoming large soft x-ray observatories.

\acknowledgments 

The authors thanks a lot to the HXI, SGD, and HXT team members of {\it Hitomi} to establish many of the key technologies supporting the FORCE mission. Works dedicated to develop Si-SOIPIX detectors are also important. Dedicated and long lasting activities of Si light-weight optics development at NASA/GSFC is also essential to this work. The science part of the mission is supported by the member of FORCE WG, as well as contributors of the ``missing BH work shop'', held at Kyoto in October 2017. Healthy discussion carried out there further strengthen the science case analysis of the mission. This work is supported by JSPS KAKENHI  15H03639, 15H02070, 16H03983, 25109004, 15H02090, 15K17648, 26610047 and 16H02170. The X-ray measurement was performed at BL20B2 in SPring-8 with the approval of the Japan Synchrotron Radiation Research Institute (JASRI) (Proposal No, 2016B1159 and 2017B1186 ).  


\bibliography{report}   
\bibliographystyle{spiebib}   

\end{document}